\documentclass{article}

\usepackage{arxiv}

\usepackage[utf8]{inputenc} 
\usepackage[T1]{fontenc}    
\usepackage{hyperref}       
\usepackage{url}            
\usepackage{booktabs}       
\usepackage{amsfonts}       
\usepackage{nicefrac}       
\usepackage{microtype}      
\usepackage{lipsum}		
\usepackage{graphicx}
\usepackage{natbib}
\usepackage{doi}
\usepackage{amssymb}
\usepackage{amsmath}
\usepackage{graphicx}
\usepackage{epsfig}

\title{The Re-Foundation of Radar Meteorology}

\author{ {Massimiliano Ignaccolo} \\
	SAS Institute, Cary, North Carolina, USA \\
	{Carlo De Michele} \\
	Politecnico di Milano, Milano, Italy\\
}

\begin{document}
\maketitle

\begin{abstract}
Since the early development of radar meteorology in the 1940s, rainfall retrieval has traditionally relied on empirical scaling relationships between radar variables and rainfall rate. Here, we revisit the premise that horizontal and differential reflectivity sufficiently constrain the rainfall rate despite the variability of the underlying drop size distribution. Motivated by evidences that this variability requires at least three degrees of freedom, we develop a nonparametric statistical retrieval algorithm that considers different combinations of drop count and distribution shape compatible with the same radar observations. The method separates the contributions of drop count and distribution shape. It then empirically estimates how the shape factors for rainfall rate and horizontal reflectivity, as well as differential reflectivity, vary with the mean and skewness of the drop diameter distribution. These relationships identify compatible combinations of drop count and shape, from which rainfall rates are calculated and their median returned as the estimate. Neither a prescribed analytical drop size distribution nor a scaling rainfall estimator or explicit classification of precipitation regimes is required. We evaluate the method using a worldwide database (149 disdrometer datasets from 76 locations), with radar observables calculated from measured drop spectra. In cross-dataset validation, the method achieves lower median dispersion of relative retrieval errors across all 16 combinations of training and validation instrument types with respect two benchmark methods consisting of the traditional scaling law relationship, and a direct nonparametric fit. These findings motivate a reassessment of the foundations of radar rainfall retrieval by incorporating the range of rainfall rates compatible with radar observation.
\end{abstract}


The retrieval of the rainfall rate from dual polarimetric radar measurements assumes that there is a functional dependence (map) $\mathcal{M}$ of the rainfall rate $R$ on the radar variables \cite{SeligaBringi1976,Ryzhkov95}. This approach extends traditional single-polarization radar methods for estimating rainfall rate from reflectivity, built on early work on raindrop size distributions by Marshall and Palmer \cite{MarshallPalmer1948}. The radar variables most commonly used are the horizontal polarization reflectivity $Z_{h}$, the ratio between the horizontal and vertical polarization reflectivities (differential reflectivity) $Z_{DR}$, and  the differential phase $K_{DP}$. Almost all conventional rainfall estimators use scaling relations involving one or two radar variables \cite{RyzhkovEtAl2022}. Other approaches use all three observables $(Z_h,Z_{DR},K_{DP})$ to retrieve DSD parameters and subsequently estimate rainfall rate \cite{BringiEtAl2002}.
In this manuscript, we focus our analysis on the commonly used couple $(Z_{h},Z_{DR})$ \cite{BringiChandrasekar2001,RyzhkovEtAl2005} although the methodology can be extended to any couple of variables. In this case, the retrieval algorithm is the so called $ZR-$relationship
\begin{equation}
\label{eq:SM}
R= \mathcal{M}(Z_{h},Z_{DR}) =a \times Z_{h}^{b} \times Z_{DR}^{c},
\end{equation}
Mathematically, the above equation implies that there is a unique value of the rainfall rate given the values of the horizontal and differential reflectivity. In practice, it means that the couple $(Z_{h},Z_{DR})$ "reasonably" fixes the value of the rainfall rate. In topological terms, it means that, for all practical purposes, the rainfall phenomenon has two degrees of freedom. 

The values of the rainfall rate and the radar variables are a function of the drop concentration per unit volume and the probability density function of drop diameters: the product of these two quantities is commonly referred to as the Drop Size Distribution (DSD): rain drops are not perfectly spherical, hence "size" is the diameter of the sphere with the same volume as the rain drop. The validity of Eq.~(\ref{eq:SM}) implies that two independent parameters are enough to quantify the statistical variability of the DSD. The most used DSD functional forms are the Gamma distribution (three parameters) \cite{Ulbrich1983} and the generalized Gamma model (four parameters) \cite{ThuraiBringi2018}. Studies such as \cite{ZhangEtAl2001,ZhangEtAl2003,VivekanandanEtAl2004}, and the references therein, discuss methods for estimating DSD parameters and their interrelationships, while \cite{BringiEtAl2003} examines their variability across climatic regimes and between convective and stratiform precipitation.

One minute disdrometer counts are commonly used to derive enough samples to estimate the triplet of parameters $(a,b,c)$. While the calculation of the rainfall rate is straightforward, the calculation of radar variables depends on the drops aspect ratio, drops orientation, drops velocity, air temperature, and angle of the incident radar beam. The introduction of the T-matrix algorithm \cite{Mishchenko1996,Mishchenko2000} has enabled a much more sophisticated and accurate estimation of the radar variables given a DSD, and thus the estimation of rainfall rate from dual polarimetric radar measurements.

In this manuscript, we build a statistical retrieval algorithm which outperforms the scaling functional relation of Eq.~(\ref{eq:SM}) even when the scaling condition is relaxed in favor of an empirically estimated one. The rationale is that no functional relationship exists, the relationship is statistical: the couple $(Z_{h},Z_{DR})$ does not "reasonably" fix the value of the rainfall rate. To a point $z$ in the $Z_{h}Z_{DR}-$space is associated not an unique value of the rainfall rate but a distribution of possible values $f_{z}(R)$. Thus, building a retrieval algorithm based on a "mapping" of $R$ in the $Z_{h}Z_{DR}-$space requires the estimation $f_{z}(R)$ for all possible points $z$, with the possible complication that $f_{z}(R)$ might also depend on the synoptic condition. Over the years, more than 100 different $ZR$-relationships have been reported in Literature \cite{IgnaccoloDeMichele2020}, and references therein, while Table 1 reports the variability observed for the 149 datasets used in our analysis. This "variability" of the ZR-relationship, in Literature, is usually solely attributed to synoptic effects. We think, that this interpretation is erroneous, and that while some synoptic effect might be present the "variability" of the ZR-relationship is an indication of the statistical nature of the relation between $R$ and the couple $(Z_{h},Z_{DR})$.

A very strong evidence supporting our thesis is that the degrees of freedom of the drop size distribution are at minimum three \cite{IgnaccoloDeMichele2022}. The Authors of \cite{IgnaccoloDeMichele2022} show that given a disdrometer drop count and not depending on synoptic conditions, the number of drops $N$, the mean $\mu$, and the skewness $\gamma$  of the probability density function of drop diameters $p(D)$ are three independent parameters that resolve $\gtrsim 95$\% (via Principal Component Analysis) of the observed variability of a disdrometer count \cite{IgnaccoloDeMichele2010,IgnaccoloDeMichele2012a,IgnaccoloDeMichele2012b}. Because the drop size distribution has a minimum of three degrees of freedom, there cannot be any functional dependence of the rainfall rate from a couple of radar variables. Any determination of the rainfall rate via a couple of radar variables has an accuracy limit which can't be surpassed due to the extra unresolved degree of freedom. Conversely, no retrieval algorithm of the rainfall rate based only on two radar variables can have an average error below a given threshold. This situation constitutes a sort of ``rainfall uncertainty principle'', similar to the one introduced by Heisenberg for conjugate variables in quantum mechanics. 

The statistical retrieval algorithm we hereby present uses the DSD parametrization of \cite{IgnaccoloDeMichele2022} to build a better retrieval model than the $ZR-$relationship. To demonstrate it, we adopt a classical data science setup.
Useful models need to reasonably fit the dataset used to build them (the training dataset) without a sensible reduction of accuracy on a dataset which was not used for training (the validation dataset). Thus, the winning model among a group of models trained on the same dataset is the one with the best validation performance. Hereby, we use $149$ disdrometer datasets from $76$ locations on Earth's surface (Fig.1). In terms of the disdrometer type, our database is composed by 27 2dvd, 36 Parsivel, 30 Joss-Waldvogel RD69, and 56 Joss-Waldvogel RD80 datasets. When a dataset is used to build a model, the remaining 148 are considered as validation. In this way, for each model (rainfall rate retrieval algorithm) we have 149 training accuracies and 22,052 ($149 \times 148$) validation accuracies.

The three models adopted in this manuscript are:
\begin{itemize}
    \item[I] Scaling Law Model (SLM): this is the scaling relationship of Eq.~(\ref{eq:SM}). Each data set is used to derive the parameters $(a, b, c)$ via least square fit in a log-log plot.
    \item[II] Direct Fit Model (DFM): each dataset is used to derive the empirical fit of the map $\mathcal{M}$ such that $R = \mathcal{M}(Z_{h},Z_{DR})$. This is a generalization of Eq.~(\ref{eq:SM}) as we relax the scaling hypothesis. We do this to further test the fact that there is no functional dependency of the rainfall rate on the couple of radar variables considered: this is so because the degrees of freedom of the DSD are at minimum 3.
    \item[III] Statistical Retrieval Model (SRM): this model uses the fact that, given a disdrometer count, for any rainfall bulk variable $X$
\begin{equation}
    \label{eq:bulkvar}
    X = I \times N \times \mathcal{F}_{X}(p(D)) =I \times N \times \mathcal{F}_{X}(\mu,\gamma)
\end{equation}
where $I$ is an instrument factor, $N$ the drop count, and $\mathcal{F}_{X}$ is a shape factor depending only on the particular shape of $p(D)$. The first equality is nothing else but a useful factorization, e.g. given a disdrometer count the shape factor for the rainfall rate is just the third moment of the pdf $p(D)$. It is important to notice that $\mathcal{F}_{X}(p(D))$ is a functional dependence in the mathematical sense: there is a unique value associated to a particular shape $p(D)$. However, it is the second equality (demonstrated in \cite{IgnaccoloDeMichele2022}) that is fundamental for building the SRM as the shape factor of any bulk variable is just a function defined in the $\mu\gamma$-plane. The couple $(\mu, \gamma)$ being the mean and skewness of $p(D)$ accounts for $\gtrsim$ 95\% (via Principal Component Analysis) of its variability \cite{IgnaccoloDeMichele2022}. Given a disdrometer dataset, the SRM uses Eq.~(\ref{eq:bulkvar}) to build the shape factors of the horizontal reflectivity $\mathcal{F}_{Z_{h}}(\mu,\gamma)$, differential reflectivity $\mathcal{F}_{Z_{DR}}(\mu,\gamma)$, and rainfall rate $\mathcal{F}_{R}(\mu,\gamma)$. Note that $Z_{DR}$ is independent from the drop count $N$ because the differential reflectivity is the ratio of two bulk variables. As matter of fact $Z_{DR}\equiv\mathcal{F}_{Z_{DR}} = \mathcal{F}_{Z_{h}}/\mathcal{F}_{Z_{v}}$. The retrieval algorithm is as follows 
\begin{itemize}
    \item  The observed value $z_{\scriptscriptstyle DR}$ of the differential reflectivity, and the shape factor $\mathcal{F}_{Z_{DR}}(\mu,\gamma)$ are used to determine the set $\mathcal{S}_{\mu\gamma}$ of couples $(\mu, \gamma)$ compatible with the observed value. 
    \item The values of each couple $(\mu, \gamma)$ in $\mathcal{S}_{\mu\gamma}$ are plugged into Eq.~(\ref{eq:bulkvar}) for $X=Z_{h}$, and retrieve the set $\mathcal{S}_{N}$ of drop counts compatible with the observed value of horizontal reflectivity $z_{\scriptscriptstyle h}$.
    \item The results of the two previous steps are used to derive the set $\mathcal{S}_{N \mu \gamma}$ of all possible triplets $(N, \mu, \gamma)$ compatible with the observed values $(z_{\scriptscriptstyle h},z_{\scriptscriptstyle DR})$. 
    \item Finally, we use Eq.~(\ref{eq:bulkvar}) with $X=R$ to obtain all possible values of the rainfall rate associated with a given dual polarimetric radar return echo: the estimate is the median.   
    \end{itemize}
\end{itemize}

To measure the train and validation accuracies of all the retrieval models, we use the bias $\mathcal{B}$ and dispersion $\mathcal{D}$. The bias of a model when predicting the values of the target variable is the median of all relative errors, while the dispersion is the interquartile range of all relative errors. The ideal model would have zero bias and zero dispersion: $(\mathcal{B},\mathcal{D}) = (0,0)$. The train accuracy of a retrieval model built on the dataset $i$ is measured by the bias and dispersion when the model fits the rainfall rates of the dataset $i$ itself. We speak of validation accuracy of a retrieval model, when we compute the couple $(\mathcal{B},\mathcal{D})$ for the predictions of the rainfall rates on a dataset $j$ different from the training dataset $i$.

Overall, to compare the accuracies of the three rainfall rate retrieval models considered in this manuscript, we compute for each model all the possible couples $(\mathcal{B}_{ij},\mathcal{D}_{ij})$ where $i,j = 1, 2, \dots, K$ ($K$ being the number of datasets considered). Training accuracies are all the couples $(\mathcal{B}_{ij},\mathcal{D}_{ij})$ where $i=j$, validation accuracies are all the couples $(\mathcal{B}_{ij},\mathcal{D}_{ij})$  where $i \neq j$. We investigate the probability density function of these accuracies to determine the best model.

\section*{Results}
Fig.~\ref{fig:results} shows how the couples ($\mathcal{B}_{ij},\mathcal{D}_{ij}$), as expressed in percent values, are distributed in the bias-dispersion plane within a regular grid of 1\% size. Every grid square is colored according to the presence in the grid of bias-dispersion couples from each of the three models adopted: white if no couples are present in the grid, light blue if only couples relative to the SRM are present, light green if only couples relative to the DFM are present, light red if only couples relative to the SLM are present, all other colors if there is a mix of couples relative to different models (in which case the color reflects the proportion of each model). The left panel shows training ($i=j$) results, and the right panel validation results ($i \ne j$). The results reported are limited to couples of datasets relative to the same type of disdrometer: one row for each type of instrument. We clearly see, both in training and in validation and not depending on the type of instrument, how the SRM model has the smaller dispersions values, and that the bias-dispersion SRM couples are clearly separated from the other two models. The SLM model, which is the current state of the art retrieval methodology, is clearly the worst of the three models: the ZR-relationship fitting a dataset is generally a poor fit when applied to a different dataset. The DFM model is better than the SLM model as the scaling condition of Eq.~({\ref{eq:SM}}) is relaxed.

In Table \ref{tab:extrainstrumenteffect}, we examine the cross-instrument effect on the validation dispersion values for all three models. We report the median, the 5-th and 95-th percentiles of validation dispersions as organized by fitting instrument (the instrument of the training dataset), fitted instrument (the instrument of the dataset used for validation), and the model. We see how:  1) The SRM is still the best of the three models 2) The SLM is still the worst of the three models 3) The performance of each model degrades when validated on a dataset relative to a different instrument. 4) The largest cross-instrument degradation occurs between the 2DVD and the remaining instruments, while the degradation within the PARSIVEL, RD69, and RD80 group are more moderate. These cross-instrument errors may reflect differences in diameter binning, sensitivity to small drops, and instrument response. Intercomparisons have documented differences among Joss--Waldvogel, PARSIVEL, and 2DVD measurements \cite{TokayEtAl2013,ThuraiEtAl2011}, and small-drop measurement limitations have also been reported for the 2DVD \cite{ThuraiBringi2018}.

\section*{Conclusions}
The Statistical Retrieval Model presented in this manuscript outperforms the standard Scaling Law retrieval Model because it does NOT attempt to fit any functional form to the relationship between the rainfall rate and the horizontal and differential reflectivity. The non functional (statistical) nature of this relation is responsible for the strong dependence of the parameters of Eq.~(\ref{eq:SM}) on the dataset adopted: there is not a fixed manifold where the triplets $(R,Z_{h},Z_{DR})$ reside. Conversely, the shape factors of rainfall bulk variables have a real functional dependence on the shape of $p(D)$ ($\mathcal{F}_{X}(p(D))$) which can be estimated using any disdrometer dataset, no matter its synoptic mixture: there is a fixed manifold where the couples $(p(D),\mathcal{F}_{X})$ reside. In practice, the estimation of the shape factors is possible because the shape of $p(D)$ is uniquely determined \cite{IgnaccoloDeMichele2022} by the mean and skewness of the distribution: $\mathcal{F}_{X}(p(D)) \rightarrow \mathcal{F}_{X}(\mu, \gamma)$.

In principle one could use the triplet of radar variables $(Z_{h},Z_{DR},K_{DP})$ to fully resolve all three degrees of freedom of the drop size distribution, but this is not the scope of this manuscript for two main reasons. 1) While given a disdrometer record, one can estimate any radar variable, the reality of radar measurements is that the three variables $(Z_{h},Z_{DR},K_{DP})$ cannot always be measured with an adequate accuracy \cite{RyzhkovEtAl2022}. Thus, there is a need to estimate the rainfall rate only with a couple of radar variables. 2) We want to prove that without imposing an analytical functional dependence to the Drop Size Distribution (Gamma distribution or generalized Gamma distribution), or adopting a scaling law among the rainfall rate and a couple of radar variables, it is possible to build an efficient and universal (reasonably good validation accuracy) rainfall retrieval algorithm.

We think that the Gamma distribution and the scaling law of Eq.~(\ref{eq:SM}) are vestiges of the historical development of radar meteorology offering an analytical treatment of the rainfall phenomenon: hence the wide utilization. However, as shown in \cite{IgnaccoloDeMichele2014,IgnaccoloDeMichele2020,IgnaccoloDeMichele2022}, and in the present manuscript they do not properly capture its statistical properties. The SRM algorithm offers a proper statistical foundation to the remote sensing retrieval of rainfall rates, making the question "which ZR relationship should I use?" a question of the past.

\section*{Data and Methods}
The datasets used in this manuscript are drawn from the
collection described in \cite{IgnaccoloDeMichele2022}. They undergo the exact same processing methodology detailed in that article and its online supplement.
\subsection*{Building a SLM}
Given a set of disdrometer observations the associated SLM model is built by calculating, via a least square fit, the parameters of Eq.~(\ref{eq:SM}) in the $(\log R, \log Z_{h}, \log Z_{DR})$ space. The predicted rainfall rate is obtained by plugging the observed values of the couple ($z_{\scriptscriptstyle h}$, $z_{\scriptscriptstyle DR}$) into Eq.~(\ref{eq:SM}).
\subsection*{Building a DFM}
Given a set of disdrometer observations the associated DFM model is built by using the Local Adaptive Fit (LAF) methodology introduced in \cite{IgnaccoloDeMichele2022} for the rainfall rate $R$ as a function of the couple $(Z_{h},Z_{DR})$.
The LAF fit adopts a grid on the space of the independent variables. Therefore, the predicted rainfall rate is the value of the mapping $\mathcal{M}(Z_{h},Z_{DR})$ for the closest grid point to the observed values of the couple ($z_{\scriptscriptstyle h}$, $z_{\scriptscriptstyle DR}$).
\subsection*{Building a SRM}
Given a set of disdrometer observations we build the associated SRM model as follows. For each disdrometer count we calculate the octuplet ($N$, $\mu$, $\gamma$, $R$, $Z_{h}$, $Z_{DR}$, $\mathcal{F}_{Z_{h}}$, $\mathcal{F}_{R}$): drop count, mean and skewness of the pdf $p(D)$, rainfall rate, horizontal reflectivity, differential reflectivity, and the rainfall rate and horizontal reflectivity shape factors (obtained by inverting Eq.~(\ref{eq:bulkvar})). The values of $Z_{h}$ and $Z_{DR}$ are obtained using the drop terminal
velocity--diameter relation of \cite{Atlas1973}, and the pytmatrix Python package \cite{Leinonen2014} with the following setting: radar wavelength$=$S-band, complex refractive indices of water at temperature=10$^{\circ}$C, drop orientation distribution=gaussian with a standard deviation of 10$^{\circ}$ degrees, drop aspect ratio functional form as in \cite{BeardChuang1987}, and all remaining settings = default settings.

The SRM model is built by using the octuplets to derive empirical curves for $Z_{DR}$, $\mathcal{F}_{Z_{h}}$, and $\mathcal{F}_{R}$ as a function of the couple $(\mu,\gamma)$. This task is done using the Local Adaptive Fit (LAF) methodology introduced in \cite{IgnaccoloDeMichele2022}. These three empirical mappings are all that is needed to retrieve the rainfall rate given an observed horizontal and differential reflectivity couple:
\begin{itemize}
    \item Given a differential reflectivity value $z_{\scriptscriptstyle DR}$, we use $Z_{DR}(\mu,\gamma)$ to obtain the set $\mathcal{S}_{\mu\gamma}$ of couples ($\mu$,$\gamma$) such that $(Z_{DR}(\mu,\gamma)-z_{\scriptscriptstyle DR})=0$. In practice this is done by retrieving the couples $(\mu_{j},\gamma_{j})$ such that the empirical fit of $Z_{DR}(\mu,\gamma)$ is within a small fraction $\epsilon$ of $z_{\scriptscriptstyle DR}$. Therefore, $\mathcal{S}_{\mu\gamma} = \{(\mu_{j},\gamma_{j}), \;j=1,\dots,M\; \textrm{such that} \;|Z_{DR}(\mu_{j},\gamma_{j})-z_{\scriptscriptstyle DR}| \leq \epsilon z_{\scriptscriptstyle DR}\}$. We set $\epsilon=$ 0.01.

    \item Given a horizontal reflectivity value $z_{\scriptscriptstyle h}$, we use Eq.~(\ref{eq:bulkvar}) to build the set $\mathcal{S}_{N}=\{N_{j},\;j=1,\dots,M\;\textrm{such that}\;N_{j}=z_{\scriptscriptstyle h}/(I\times\mathcal{F}_{Z_{h}}(\mu_{j},\gamma_{j}))\;\forall\;(\mu_{j},\gamma_{j})\in\mathcal{S}_{\mu\gamma}\}$.

    \item We combine the sets $\mathcal{S}_{\mu\gamma}$ and $\mathcal{S}_{N}$ to obtain the set $\mathcal{S}$ of all the triplets $(N_{j},\mu_{j},\gamma_{j})$ ($j=1,\dots,M$) which are compatible with the observed values of horizontal reflectivity $z_{\scriptscriptstyle h}$ and differential reflectivity $z_{\scriptscriptstyle DR}$.

    \item We use Eq.~(\ref{eq:bulkvar}) to obtain $M$ possible values of rainfall rate, and take the median value $r$ to be the predicted rainfall rate given ($z_{\scriptscriptstyle h}$, $z_{\scriptscriptstyle DR}$).
\end{itemize}
The SRM model finds the values of drop count and the shapes of the drop size distribution (and thus the possible rainfall rate values) that are statistically compatible with the observed reflectivities: the final predicted rate being the median value of the compatible rainfall rates. Note that $M$ is not a fixed constant as it depends on the particular value of the couple ($z_{\scriptscriptstyle h}$, $z_{\scriptscriptstyle DR}$).

\subsection*{Assessing model accuracy}
A model $\mathcal{M}$ built on the dataset $i$ and used to fit the dataset $j$ has a bias $\mathcal{B}_{ij}$ and dispersion $\mathcal{D}_{ij}$ which are defined, respectively, as the median and the interquartile range of all relative errors. The training accuracy of a model is defined by the couple $(\mathcal{B}_{ii},\mathcal{D}_{ii})$. The case $i\ne j$ constitutes a validation accuracy. For each rainfall rate retrieval algorithm we can build 149 models (one for each dataset) leading to 149 possible values of training accuracy and $149\times148$$=$$22,052$ possible values of validation accuracies.

\begin{figure*}[h]
    \centering   
    \epsfig{file=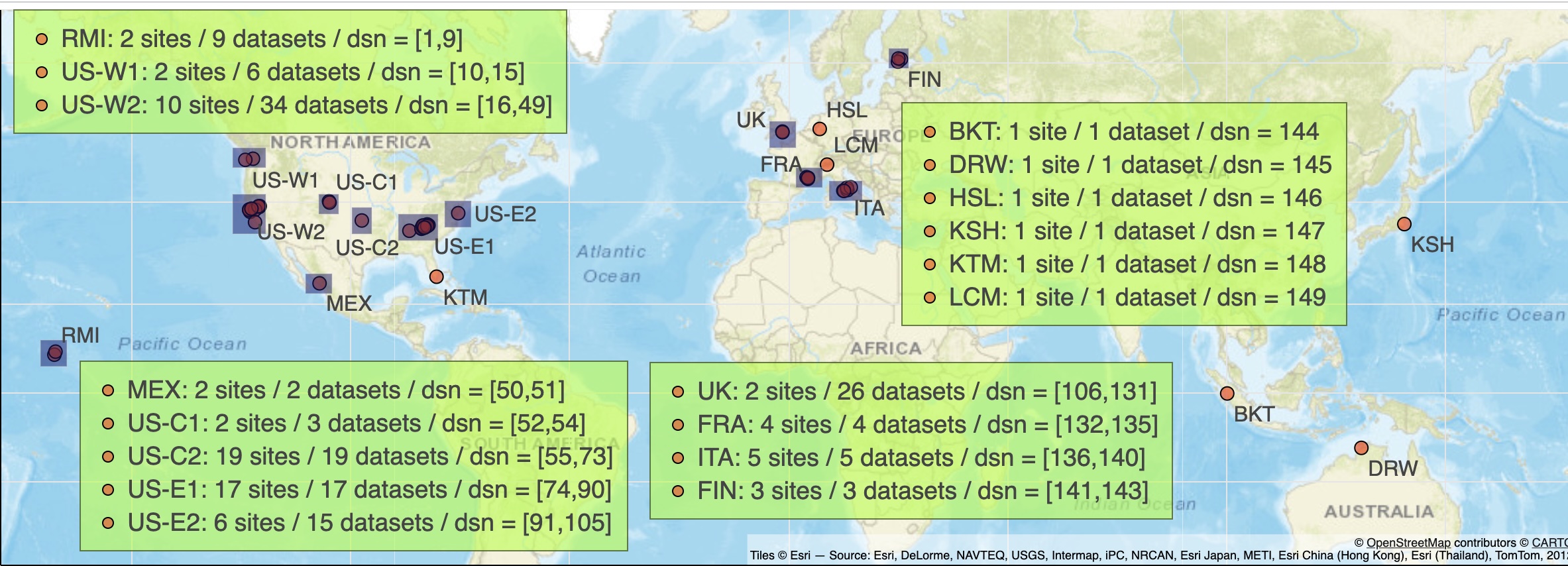, width=1.0\linewidth, angle=0}
    \caption{Sites’ locations: orange circles. Given the map resolution, many locations cannot be fully resolved: we
use a "group" label and shadow the sites’ locations with blue rectangles. For each label in the figure, we report
1) the number of distinct sites associated with it, 2) the number of data sets derived from the sites, 3) the data sets number in our database.}
    \label{fig:1}
\end{figure*}

\begin{table}[htbp]
  \centering
  \setlength{\tabcolsep}{3pt}     
  \label{tab:nometabella}
  \begin{tabular}{|l|c|c|c|c|c|c|c|}
    \toprule
    Parameter & Mean & Std Dev & Min & Q1 & Median & Q3 & Max \\
    \midrule
    a & 0.0235 & 0.0064 & 0.0091 & 0.02078 & 0.0259 & 0.0284 & 0.0328 \\
    b & 0.807 & 0.053 & 0.672 & 0.782 & 0.815 & 0.845 & 0.939 \\
    c & -6.4 & 2.4 & -11.6 & -8.4 & -6.4 & -4.4 & -2.1 \\
    \bottomrule
  \end{tabular}
  \caption{The mean, standard deviation, minimum, first quartile, median, third quartile, and maximum value for the parameters of the scaling equation Eq.~(\ref{eq:SM}) for the 149 datasets considered in the manuscript.}
\end{table}

\begin{table}[htbp]
  \centering
  \setlength{\tabcolsep}{3pt}     
  \begin{tabular}{|l|c|c|c|c|c|c|}
    \toprule
    Fitting & Fitted & Model & 50\% $\mathcal{D}_{ij}$ & 5\% $\mathcal{D}_{ij}$ & 95\% $\mathcal{D}_{ij}$ & Samples\\ 
    \midrule
    2DVD & 2DVD & SRM & 0.124 & 0.088 & 0.158 & 695\\
     & & DFM & 0.212 & 0.165 & 0.331 & 702\\
     & & SLM & 0.259 & 0.160 & 0.325 & 702\\
    \cline{2-7}
     & PARSIVEL & SRM & 0.247 & 0.176 & 0.360 & 950\\
     &  & DFM & 0.431 & 0.335 & 0.658 & 958\\
     &  & SLM & 0.477 & 0.339 & 0.613 & 972\\
    \cline{2-7}
     & RD69 & SRM & 0.353 & 0.264 & 0.494 & 781\\
     &  & DFM & 0.413 & 0.325 & 0.642 & 809\\
     &  & SLM & 0.619 & 0.311 & 0.815 & 810\\
    \cline{2-7}
     & RD80 & SRM & 0.354 & 0.252 & 0.528 & 1475\\
     &  & DFM & 0.405 & 0.310 & 0.685 & 1512\\
     &  & SLM & 0.542 & 0.306 & 0.775 & 1512\\
    \hline
   PARSIVEL & 2DVD & SRM & 0.147 & 0.098 & 0.216 & 972\\
     & & DFM & 0.217 & 0.161 & 0.306 & 972\\
     & & SLM & 0.297 & 0.212 & 0.404 & 972\\
    \cline{2-7}
     & PARSIVEL & SRM & 0.127 & 0.102 & 0.174 & 1260\\
     &  & DFM & 0.254 & 0.204 & 0.340 & 1260\\
     &  & SLM & 0.352 & 0.273 & 0.457 & 1260\\
    \cline{2-7}
     & RD69 & SRM & 0.161 & 0.129 & 0.206 & 1078\\
     &  & DFM & 0.227 & 0.194 & 0.258 & 1080\\
     &  & SLM & 0.380 & 0.279 & 0.615 & 1080\\
    \cline{2-7}
     & RD80 & SRM & 0.154 & 0.118 & 0.204 & 2015\\
     &  & DFM & 0.229 & 0.187 & 0.310 & 2014\\
     &  & SLM & 0.372 & 0.259 & 0.614 & 2016\\
    \hline
   RD69 & 2DVD & SRM & 0.200 & 0.140 & 0.280 & 810\\
     & & DFM & 0.270 & 0.196 & 0.346 & 810\\
     & & SLM & 0.406 & 0.267 & 0.458 & 810\\
    \cline{2-7}
     & PARSIVEL & SRM & 0.168 & 0.117 & 0.250 & 1078\\
     &  & DFM & 0.281 & 0.224 & 0.355 & 1080\\
     &  & SLM & 0.404 & 0.320 & 0.485 & 1080\\
    \cline{2-7}
     & RD69 & SRM & 0.148 & 0.122 & 0.192 & 870\\
     &  & DFM & 0.229 & 0.198 & 0.266 & 870\\
     &  & SLM & 0.356 & 0.303 & 0.493 & 870\\
    \cline{2-7}
     & RD80 & SRM & 0.151 & 0.117 & 0.295 & 1680\\
     &  & DFM & 0.229 & 0.198 & 0.287 & 1680\\
     &  & SLM & 0.391 & 0.311 & 0.485 & 1680\\
    \hline
   RD80 & 2DVD & SRM & 0.192 & 0.123 & 0.295 & 1511\\
     & & DFM & 0.259 & 0.178 & 0.356 & 1512\\
     & & SLM & 0.352 & 0.233 & 0.473 & 1512\\
    \cline{2-7}
     & PARSIVEL & SRM & 0.157 & 0.112 & 0.251 & 1975\\
     &  & DFM & 0.300 & 0.222 & 0.403 & 2008\\
     &  & SLM & 0.391 & 0.303 & 0.511 & 2016\\
    \cline{2-7}
     & RD69 & SRM & 0.144 & 0.113 & 0.188 & 1679\\
     &  & DFM & 0.230 & 0.191 & 0.305 & 1679\\
     &  & SLM & 0.348 & 0.288 & 0.450 & 1680\\
    \cline{2-7}
     & RD80 & SRM & 0.143 & 0.109 & 0.188 & 3080\\
     &  & DFM & 0.233 & 0.191 & 0.322 & 3080\\
     &  & SLM & 0.355 & 0.270 & 0.460 & 3080\\
    \bottomrule
  \end{tabular}
  \caption{The median, 5-th quantile, and 95-th quantile of the all validation dispersions. Values are organized by instrumentation of the training and validation datasets and by retrieval model. The Samples column report the number of training-validation couples for each instrument-model combination.}
  \label{tab:extrainstrumenteffect}
\end{table}

\begin{figure}[h]
    \centering   
    \epsfig{file=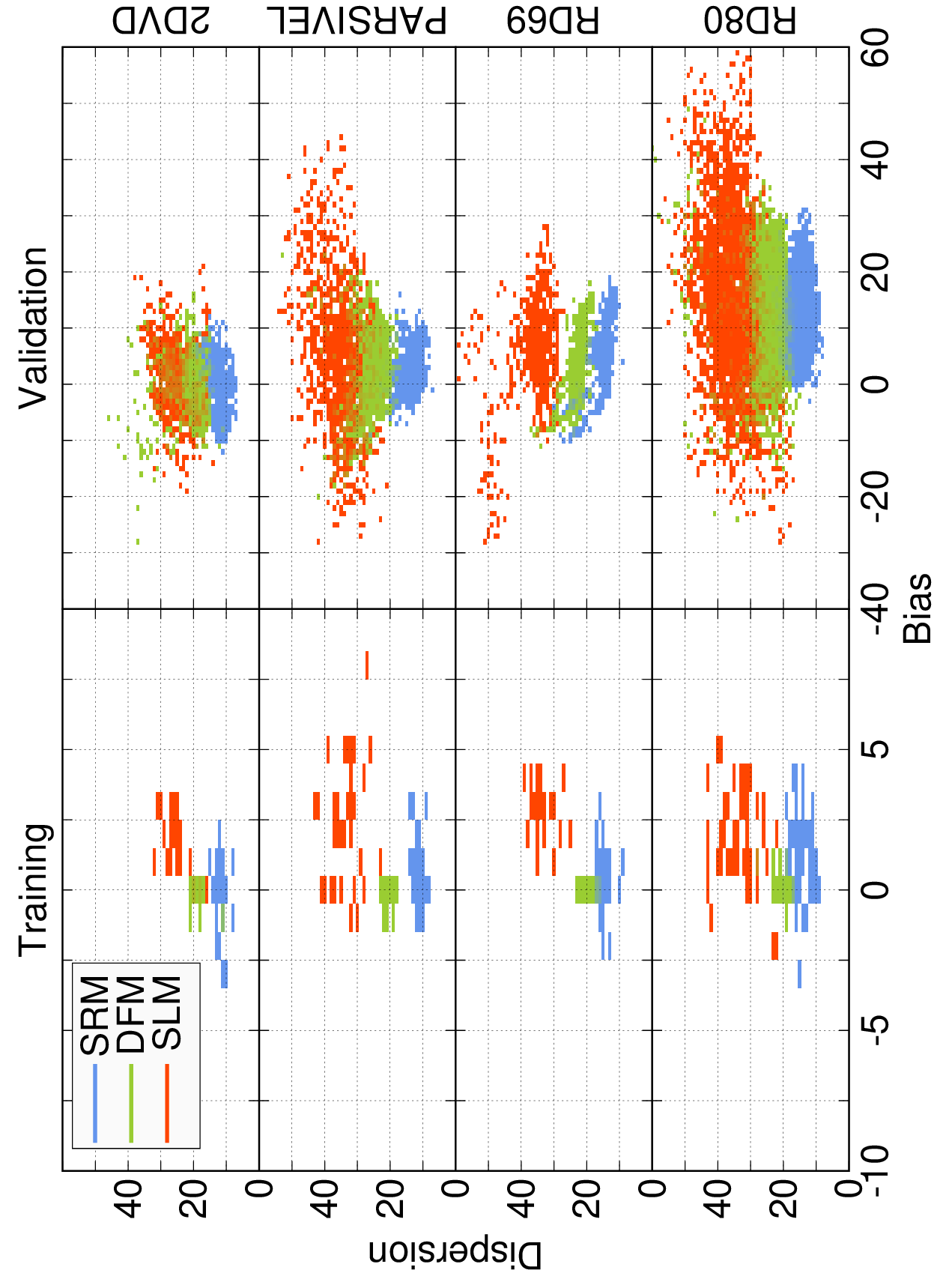, width=0.75\linewidth, angle=-90}
    \caption{The distribution of the bias-validation couples $(\mathcal{B}_{ij},\mathcal{D}_{ij})$ for all three models considered. 
    Color reflect grid occupancy according to model. Left column refers to training, and right to validation. Different rows segement the results by datasets instrumentation.}
    \label{fig:results}
\end{figure}


\section*{Acknowledgment}
We wish to sincerely thank colleagues and institutions who have generously provided disdrometer data: 1) Dr. C. R. Williams, the National Oceanic and Atmospheric Administration (NOAA) and Physical Sciences Laboratory (PSL) for the public availability of the datasets recorded at Alta (16), Bodega Bay (DSN 17), Erie (DSN 52), Cazadero (DSN 27 to DSN 38), Colfax (DSN 18 to DSN 26), Darwin (DSN 145), Davis (DSN 39), Estacion Obispo (DSN 50 and 51), Fort Ross (DSN 40), Grass Valley (DSN 41), Point Sur (DSN 48 to DSN 49), Rocky Mtn. Arsenal (DSN 54), Spanaway (DSN 10), Westport (DSN 11 to DSN 15); 2) Dr.  Ali Tokay and National Aeronautics and Space Administration (NASA) for the public availability of the following datasets: Wallops Precipitation Research Facility (DSN 91 to DSN 105), IFloodS Alabama disdrometer data (DSN 78 and 79), Crystal Face Florida disdrometer data (DSN 148), Kwajalein Disdrometer Data (DSN 1 to DSN 9), LPVex Disdrometer Data (DSN 141 to DSN 143), and as part of the  Global Precipitation Measurement (GPM) mission Ground Validation campaigns: HYMEX (DSN 133 to DSN 140), IPHEX (DSN 74 to DSN 77, and DSN 80 to DSN 90), MC3E (DSN 55 to DSN 73); 3) the National Institute of Information and Communications Technology, Japan for the Kashima data set (DSN 147); 4) the Institute of Observational Research for Global Change together with the Japan Agency for Marine-Earth Science and Technology for the Bukit Koto Tabang data set (DSN 144); 5) Dr. Dan Brawn for the Hassel data set (DSN 146); 6) Dr. Martin Hagen of the Institut für Physik der Atmosphäre Deutsches Zentrum für Luft- und Raumfahrt, Wessling, Germany for the Macunaga data set (DSN 149); 7) Dr. Charles Wrench and the British Atmospheric Data Centre (BADC) for Chilbolton (DSN 106 to DSN 121) and Sparsholt (DSN 122 to DSN 131) data archives.

\clearpage
\bibliographystyle{unsrt}
\bibliography{reference}








\end{document}